# Fractal in the statistics of Goldbach partition

Wang Liang[1]*, Huang Yan[2] , Dai Zhi-cheng[1]

[1,2] (Department of Control Science and Control Engineering, Huazhong University of Science and Technology, WuHan, 430074, P.R.China)

[2](Department of Mathematics, Huazhong University of Science and Technology, WuHan 430074, P.R.China)

**[Abstract]** Some interesting chaos phenomena have been found in the difference of prime numbers. Here we discuss a theme about the sum of two prime numbers, Goldbach conjecture. This conjecture states that any even number could be expressed as the sum of two prime numbers. Goldbach partition r(n) is the number of representations of an even number n as the sum of two primes. This paper analyzes the statistics of series r(n) (n=4,6,8,……). The familiar 3 period oscillations in histogram of difference of consecutive primes appear in r(n).We also find r(n) series could be divided into different levels period oscillation series. The series in the same or different levels are all very similar, which presents the obvious fractal phenomenon. Moreover, symmetry between the statistics figure of sum and difference of two prime numbers are also described. We find the estimate of Hardy-Littlewood could precisely depict these phenomena. A rough analyzing for periodic behavior of r(n) is given by symbolic dynamics theory at last.

## 1 Introduction

Recently, the interest in prime numbers received a new impulse. Prime number sequence is found in many unrelated areas. Examples range from the periodic orbits of a system in quantum chaos to the life cycles of species [1-3].

On the other hand, some new methods in chaos and statistic theory are also applied to study the primes. Normally, the differences of primes are used. The famous 3 period oscillation was found in statistics histogram of difference of consecutive primes many years ago[4]. It also happens in the histogram of increment (difference of difference) of consecutive prime [5][6]. The difference of primes in Dirichlet classification appears the same special periodic behavior [7]. Paper [8] shows its relation with the celebrated Sierpinski fractal. The interesting “Jump Champion” problem also belongs to this kind research [9].

Goldbach conjecture is a theme about the sum of primes. It states that every even integer > 2 can be expressed as the sum of two primes. The proof remains an unsolved problem since Goldbach first wrote the conjecture in a letter to Euler in 1792. However, significant progress has been made in recent years.

In 1855, Desboves verified Goldbach Conjecture for n < 10000. The recent record is $n < 2\times 10^{17}$ [10] .No counter-example has been found to date. Till now, the best theory result is J. R. Chen's 1966 theorem that every sufficiently large integer is the sum of a prime and the product of at most two primes [11]. Some new methods in other area are also applied to study this conjecture. For example, paper [12] gives it a novel discussion by small world network theory.

This paper is mainly concerned with the statistic character of Goldbach partition. Goldbach partition r(n) is the number of representations of an even number n as the sum of two primes. For example:

$$100 = 3 + 97 = 11 + 89 = 17 + 83 = 29 + 71 = 41 + 59 = 47 + 53$$

So r(100)=6.Here ‘3+97’ and ‘97+3’ are regarded as the same representation.

We find the interesting fractal character in the detailed structure of series r(n). The subtle symmetry also appears when comparing the r(n) with the statistics of difference of two prime numbers. In the second section of this paper, we just introduce these experimental results. Then a theory discussion is given in the third section. Here we build the relation between a simple unimodal mappings and r(n) by symbolic dynamics theory. A referenced Lyapunov exponent of series r(n) is also obtained through this relation. Conclusions are summarized in the last section.

* Corresponding author: guoypm@hust.edu.cn

## 2 Experimental results

### 2.1 Three period oscillations

Here r(n) is defined as the Goldbach partition, n is even number. We could calculate the Goldbach partition of consecutive even numbers and plot it in a figure (Fig 1).

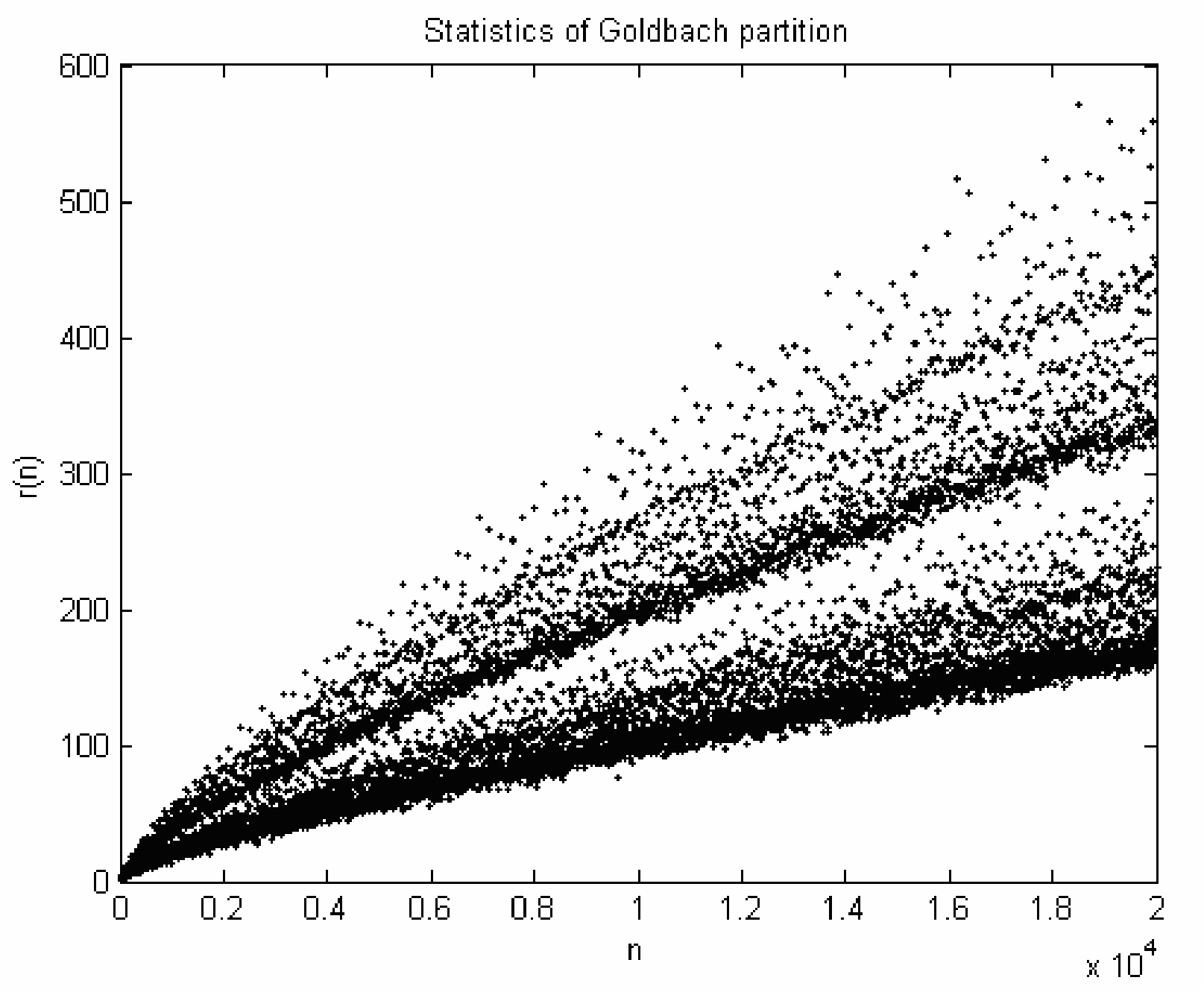

Fig 1. Figure of series $r(n), 6 < n < 2\times 10^4$

This figure is sometimes called the "Goldbach Comet". It appears in paper [13] first. Here Henry Fliegel of Aerospace Corporation, and Douglas Robertson of National Geodesic Survey in 1989 computed r(n) for $n < 1\times 10^5$ and plotted it. What they found was that the graph had two distinct bands and a sharp lower edge, resembling the appearance of a comet. It shows that generally the number of distinct representations increases with increasing n. An asymptotic approach appears to provide a possible avenue for success in proving out the Goldbach Conjecture. Hardy and Littlewood had derived an estimate for r(n) in 1923 [14]:

$$r(2n) \sim \frac{2nC}{(\log n)^2} \prod_{p>2, p|n} \frac{p-1}{p-2}$$

Here, $C = \prod_{p>2} (1 - \frac{1}{(p-1)^2}) = 0.66016\cdots\cdots$

While Goldbach comet supports the Goldbach conjecture, it of course does not prove anything. It is opinions of many researchers. So there are still no many researches for it. Some recent work for its lower/upper bounds could be found in [15][16].For us, the obvious band structure in the figure is more attractive. The figure is clearly divided into upper part and lower part, which have some similarities. Some well-regulated subtle structure also appears in these two parts. Maybe there is something linking the fractal and chaos. This is our interests.

Just like the research about the statistics of difference of consecutive prime number (Fig 2.a), we could also select some segments of r(n) and plot them in the form of histogram(Fig 2.b, c ).

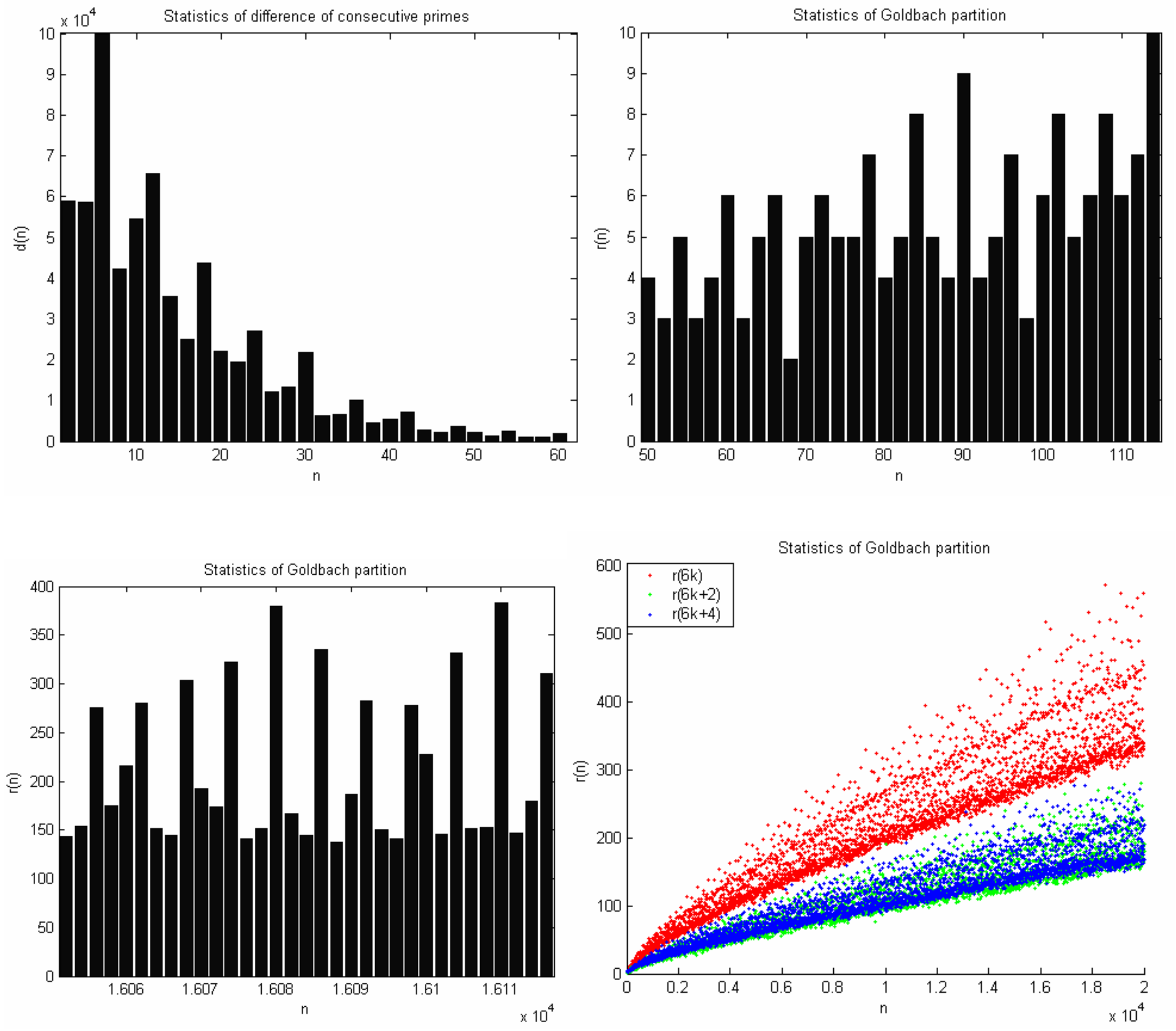

Fig 2. (a) Statistics histogram of differences for the first 10 million of consecutive prime numbers. (b) Histogram of Goldbach partitions r(n), 50<n<114.(c) Histogram of r(n), a segment of n>15000.(d) r(n) in different colors

We find the 3 period oscillations in histogram of difference of consecutive primes also appear in the histogram of r(n). This phenomenon could be described as:

$$\begin{matrix} r(6k) > r(6k+2) \\ r(6k) > r(6k+4) \end{matrix}, \text{ k is natural number, k >25}$$

To see it more clearly, we plot the figure of $r(n), n = 6k, 6k+2, 6k+4$ in different color (Fig 2.d). We found series r(6k) is clearly bigger than others. r(6k+2) and r(6k+4) are almost mixed together, but the low bound is r(6k+2) mostly. The detailed numerical calculating of $r(6k)-r(6k+2), r(6k)-r(6k+4), 150<6k<5\times10^4$ also proves it. So we guess this character may be scale invariant. The oscillation of statistics of difference of consecutive primes still can't be explained strictly, so does this sum of primes.

Here is a probability explanation for this band structure of series r(n). Consider

$$6k\equiv 0\equiv 1+5 \pmod 6, 6k+2\equiv 2\equiv 1+1 \pmod 6, 6k+4\equiv 4\equiv 5+5 \pmod 6$$

So even numbers in 6k can take from all the primes in 6k+1 and 6k+5, but even numbers in 6k+2, 6k+4 can only take primes from 6k+1 or 6k+5 exclusively. So generally their Goldbach partitions $r(6k)>r(6k+2), r(6k)>r(6k+4)$. The multiplier $\prod_{p>2,p|n}\frac{p-1}{p-2}$ in the estimate of Hardy-Littlewood could also give a rough explanation. Here we can't deduce $r(6k)>r(6k+2), r(6k)>r(6k+4)$ by Hardy-Littlewood conjecture because we didn't know the detailed distribution of primes. But these plausible explains seem very precise.

To make a comparison, we also plot the estimate of Hardy-Littlewood in the form of dot figure and histogram (Fig 3.a, b).Here we denote: $h(2n)=\frac{2nC}{(\log n)^2}\prod_{p>2,p|n}\frac{p-1}{p-2}$ .we find the h(n) has the same periodic character as r(n).

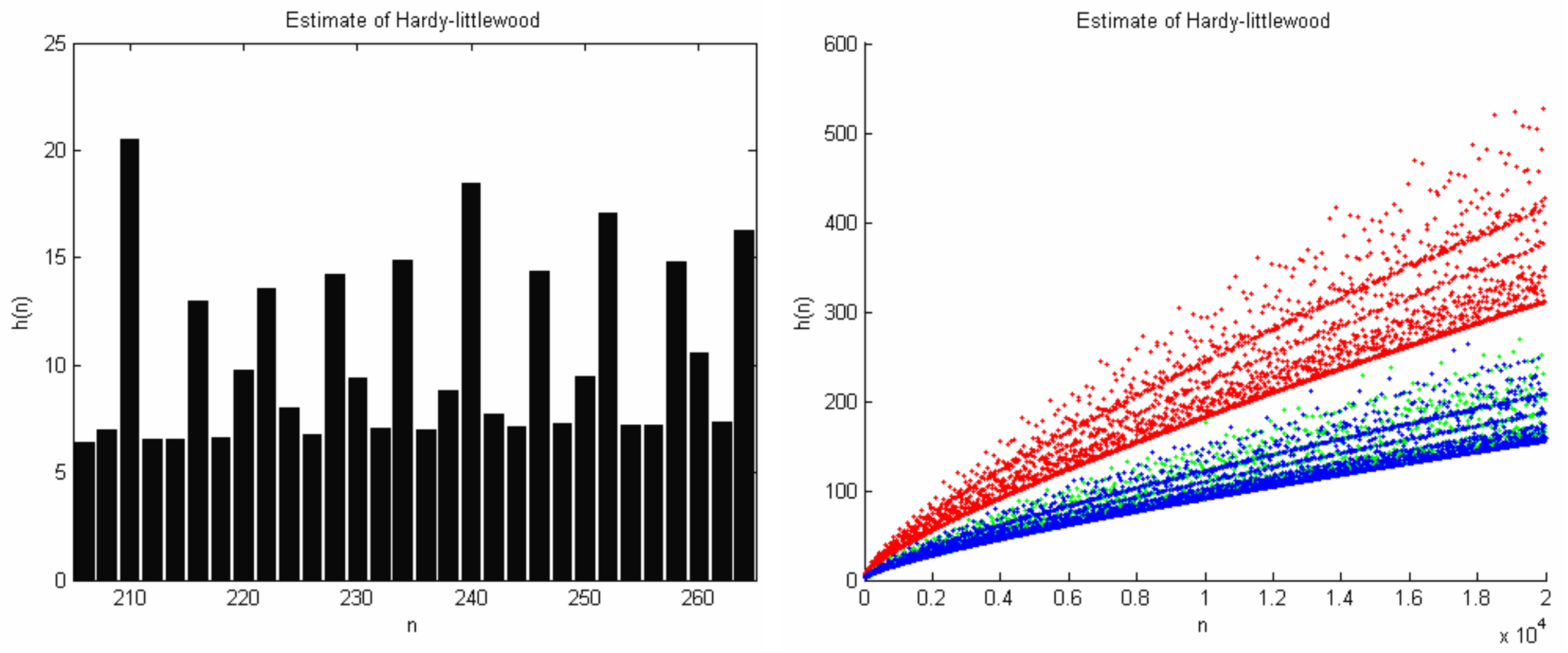

Fig 3 (a) histogram of h(n), (b) dot figure of h(n)

The histogram only contains a small segments of r(n) and describes the detailed structure. The dot figure of r(n) depicts the general structure of r(n) .So this finding could be described as:

**Observation 1:**

**Histogram form: The r(n) is a 3 period oscillations series.**

**Dot figure form: Figure of r(n) has three parts, a upper parts and two lower parts mixed together. We could**

**also say figure of r(n) is divided into 2 independent parts.**

**2 .2 Fractal**

To make the description more clear, We rewritten series r(n) as follows:

$$r_1 = r(6), r_2 = r(8), r_3 = r(10), \cdots\cdots, r_n = r(2 \times n + 4)$$

Then we disjoint the r(n) into three new series, which are marked as $A_1, A_2, A_3$ respectively.

$$r : r_1, r_2, r_3, \cdots, r_n$$
$$A_1 : r_1, r_4, r_7, \cdots r_{3i}$$
$$A_1 : r_2, r_5, r_8, \cdots r_{3i+1}$$
$$A_1 : r_3, r_6, r_9, \cdots r_{3i+2}, i = 1,2,3,4,5\cdots$$

It expresses the coloring method in paragraph 2.1. We call it "series disjoint method" with interval 3, which select element in the same place of every period and construct 3 new series. In fact, it's a common method in some chaos time series analyzing technologies like C-C method.

To study their detailed structure, we could also plot the $A_1, A_2, A_3$ in histogram form. A random segment is selected from series $A_1, A_3$ (Fig 4).

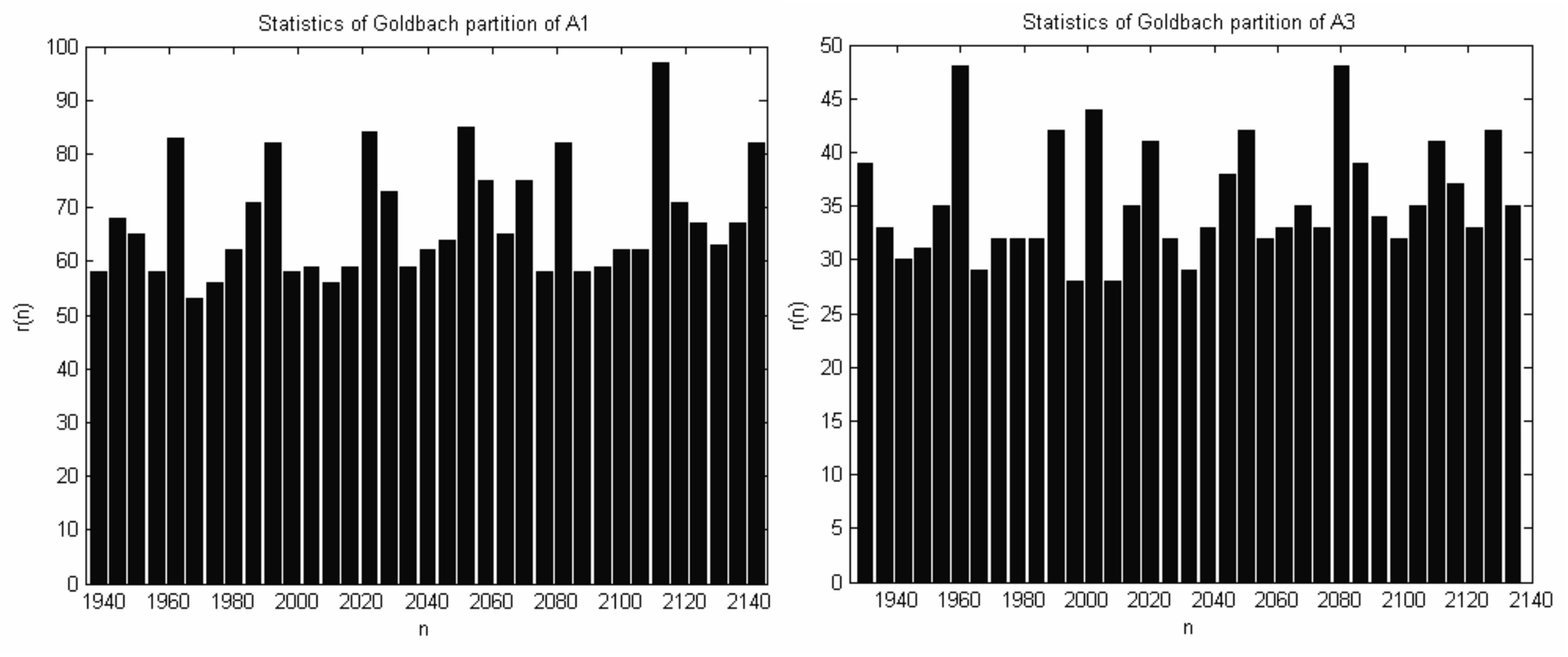

Fig 4. Histogram of $A_1, A_3$

Interesting thing appears. We found these series all have 5 period oscillation with few exceptions We verify this character before $n < 5 \times 10^4$. In $A_1$, there is 69 exception , accounting for 4%. Exception in $A_2$ is 6%, $A_3$ is 8%. The exceptions in the series are equably distributed. This oscillation is very precise.

Here we could also divide $A_1$ into 5 new series $B_1, B_2, B_3, B_4, B_5$ by "series disjoint method". The maximal series is plotted by red color, other four by blue. $A_2$ is also divided into 5 new series and plot in the same method (Fig 5.a). The figure of $A_1, A_2$ all have a clear upper part and a lower part. We could found there is some blue dots in the maximal sub-series of $A_1$. It's just the exception periods. The Fig 5.b shows the maximal sub-series of $A_1$ and exception points alone.

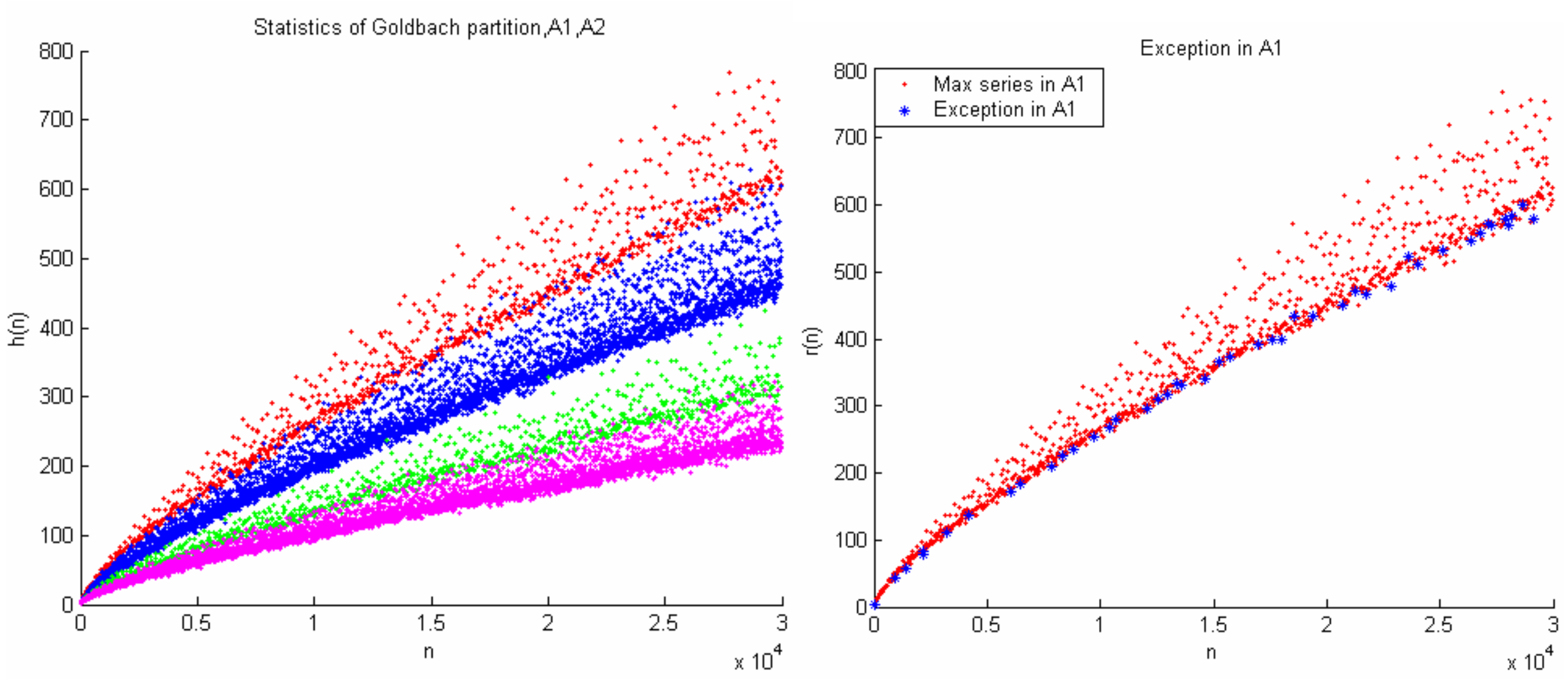

Fig 5. (a) Figure of series $A_1, A_2$ (b) Exception dot in $A_1$.

We found in A1:

$$\begin{array}{l} r(6\times 5k) > r(6\times(5k+i)) \\ i = 1,2,3,4 \end{array}$$ , k is natural number.

In A2:

$$\begin{array}{l} r(6\times(5k+3)+2) > r(6\times(5k+i)+2) \\ i = 0,1,2,4 \end{array}$$ , k is natural number.

In A3:

$$\begin{array}{l} r(6\times(5k+1)+4) > r(6\times(5k+i)+4) \\ i = 0,2,3,4 \end{array}$$ , k is natural number.

We also divide the h(n) into three new series and plot them in two forms(Fig 6). Few exception periods also exist in h(n).

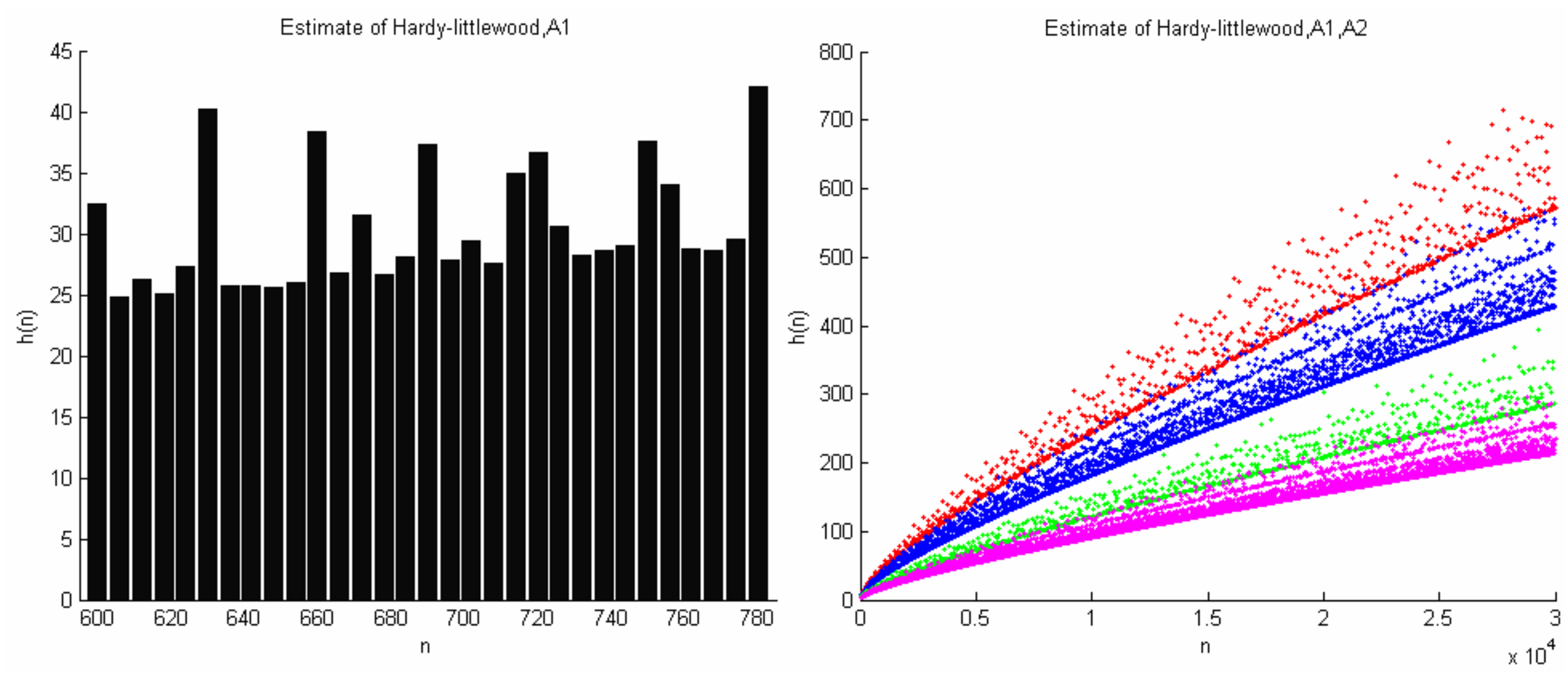

Fig.6 Histogram and figure of h(n)

We have:

**Observation 2:**

**Histogram form: The r(n) is a 3*5=15 period oscillations series**, which is embedded with 3 period oscillations.

**Dot figure form:** Figure of r(n) has three parts, a upper parts and two lower parts mixed together. **Every of these three parts also comprises 5 parts, an upper parts and four lower parts mixed together. We could also say Figure of r(n) is divided into 4 'independent' parts.**

We go ahead. If looking into the detailed structure of new series $B_1$ divided from A1, we find it is oscillation series with period 7. Other four series also have 7 period oscillations. We plot one of them in the form of histogram (Fig 7).

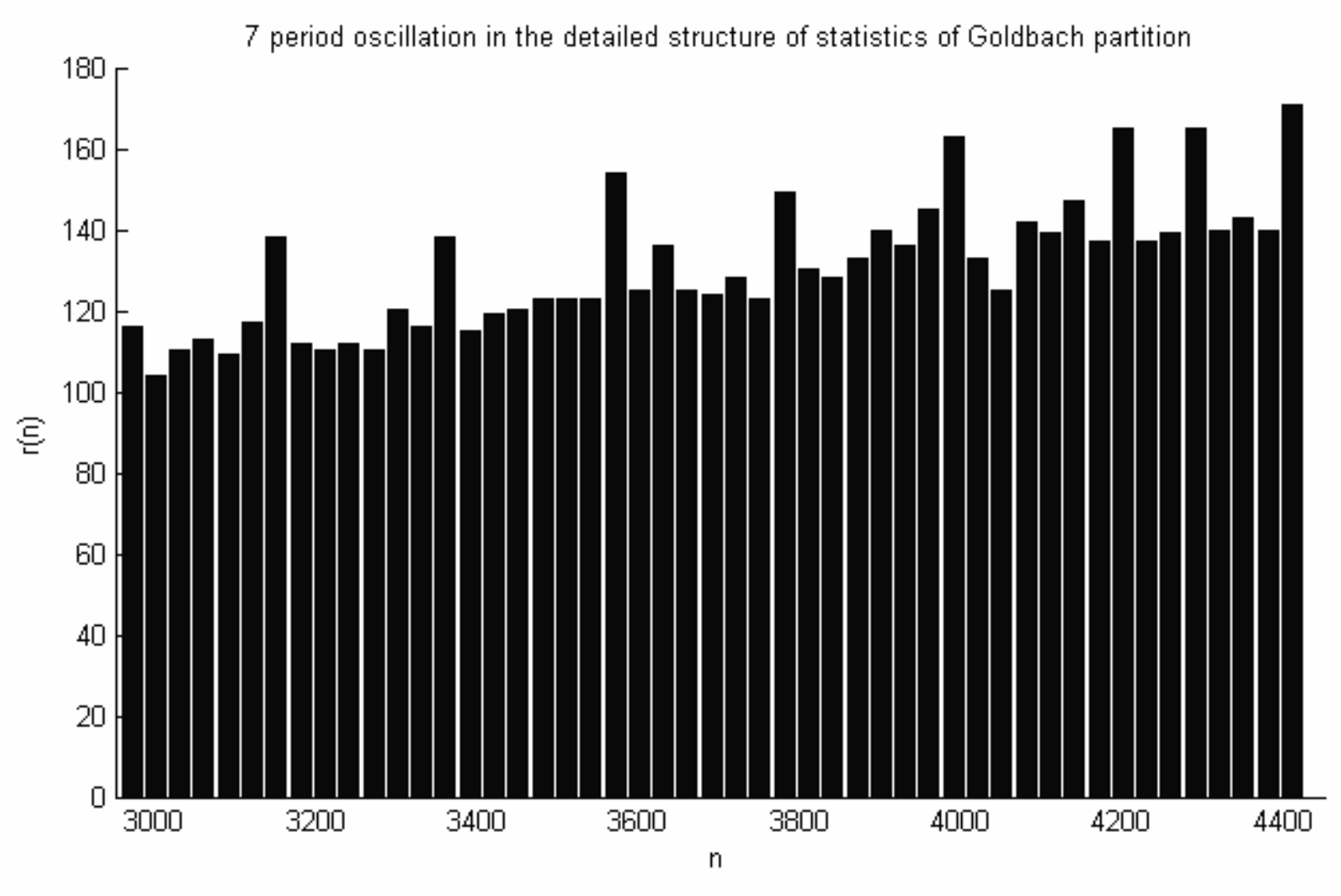

Fig 7. Histogram of B1

Then we divide $B_1$ into 7 new series by "series disjoint method". We find there is also a maximal sub series in $B_1$. The others are mixed together. The exception periods accounts 1% in this figure. This phenomenon also happens in $A_2, A_3$. The exception in other series also doesn't exceed 10%.

We plot the maximal sub-series by red color and other series of $B_1$ by blue. Because $B_1$ is the maximal sub series of $A_1$. So we select one lower sub-series of $A_1$ and divide it into 7 new series. Its maximal sub series is also plotted by green color and other 6 series by pink. We also do the same thing for $A_2$ and plot it in the same map (Fig.8.a). h(n) is also plotted in the same way (Fig.8.b).

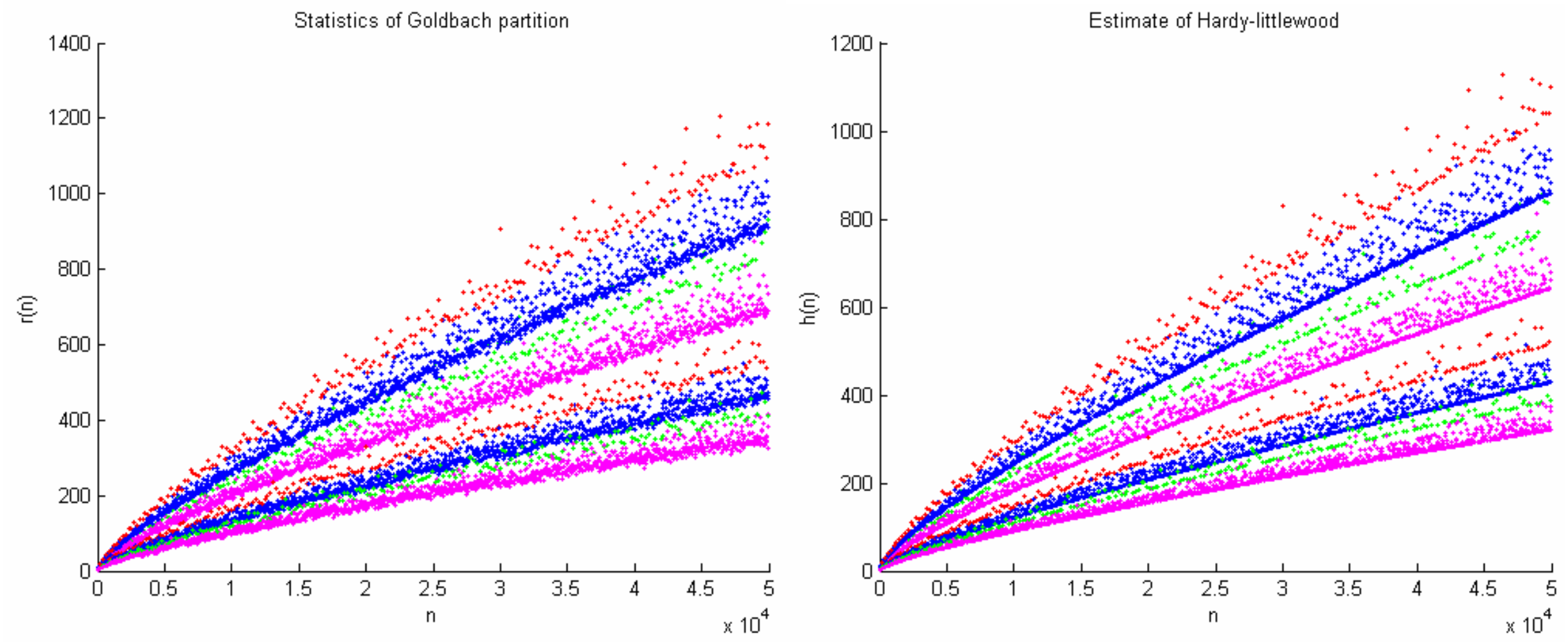

Fig 8. (a). Figure of $A_1, A_2$ .(b). Figure of h(n)

Till now, we have:

**Observation 3:**

**Histogram form: The r(n) is a 3*5*7=105 period oscillations series** embedded with 3*5=15 period oscillations. These 15 period oscillations are also embedded with 3 period oscillations.

**Dot figure form:** Figure of r(n) has three parts, a upper parts and two lower parts mixed together. Every of these three parts comprises of 5 parts, an upper parts and four lower parts mixed together. **Every of these 5 parts also has 7 parts, an upper part and six lower parts mixed together. We could also say Figure of r(n) is divided into 8 'independent' parts.**

Just as we wish, all 7 piece of sub-series divided from $B_1$ is 11 periodic series. 13 periods are also found in the more detailed structure. So we guess all prime number periods will appear if getting enough data of r(n). In fact, if we want to find the 23 period oscillation, the series r(n) should be divided into $3\times5\times7\times9\times11\times13\times17\times19\approx5\times10^6$ piece of new series. We may also need at least 200 data in one series. So all the Goldbach partition of $n, n<2\times10^9$ should be calculated. It's a difficult mission for personal computer, but maybe a good research topic for Grid Computing.

We stop here and conclude the operation above as follows:

(1) By "series disjoint method", r(n) could be divide into 3 piece of new independent series , which are all 5

periods.

(2) Each of these 3 series could also be divided into 5 new independent series. These $3\times 5=15$ series are all 7 periods.

(3) Every of these 15 new series is divided into 7 new independent series. These $3\times 5\times 7=105$ series all have 11 periods.

If combining these three steps together, we could say that the r(n) is divided into $3\times 5\times 7=105$ new series by "series disjoint method" and all these series will have 11 periods. Continuing this operation, we guess bigger prime number period will appear in the more detailed structure of r(n). We summarize this operation in a precise form and get conjecture:

**Conjecture 1:**

**Let** $P_m = 3\times 5\times 7\times\cdots\times p_m = \prod_{i=2}^{m} p_i$ **,** $p_i$ **is the ith prime number,** $m = 2,3,4,\cdots$

**The series** $r(n), n\to\infty$ **could be divided into** $P_m$ **pieces of new independent series by "series disjoint method". These new series all have the similar** $p_{m+1}$ **period oscillations with few exceptions.**

This also means the figure of r(n) could be divided into $2^{m-1}$ 'independent' parts. So there is no a thinnest bands acting as the lower/upper bounds of r(n) figure because we could always find the thinner bands. Under this meaning, there are no exact expressions for the bounding curves of figure r(n). It's just like that famous question "How long is coast of Britain"[17]. Moreover, if we map a small segment of r(n) to a vertical line of x axis, this section of r(n) will be very similar with the 3 division Cantor Set. So an interesting guess is:

**"Goldbach Comet"="Britain coast" + "Cantor set".** We will discuss the bounds and sections of r(n) in another paper.

If regarding r(n) as a time series, we could say r(n) is $P_m$ period oscillations series embed with $P_{m-1}, P_{m-2}, \cdots, P_2$ period oscillations. Similar periodic behavior also happens in some actual communication signals. It's said that just such signals give Mandelbrot inspiration to create the concept of fractal. This kind of signal is also the standard research object of a new technology "Wavelets Analysis".

We mark the series $A_1, A_2, A_3$ as the first level series, and $B_1,\cdots,B_5,\cdots$ as the second level series, etc. The figures of the series in different levels all have an upper parts and lower part. A maximal sub-series is clearly bigger than others and the others sub-series are mixed together. .All these series also have the similar asymptotic function. So we could say series r(n) has self-similarity character. Part of this phenomenon could be described as:

**Conjecture 2:**

**Let** $P_n = 2\times 3\times 5\times 7\times\cdots\times p_n = \prod_{i=1}^{n} p_i$ **,** $p_i$ **is the ith prime number,** $n = 1,2,3,\cdots,\infty$

**Then:**

$r(P_n\times k) > r(P_n\times k + i), i = 2,4,6,\cdots,P_n - 2$, **k is natural number. Few exceptions are distributed in the whole series.**

There are self-similarity and infinite hierarchies in "Goldbach Comet". So our finding could be described in one word that the "Goldbach Comet" is a typical fractal figure.

**2.3 Symmetry between sum and difference of primes**

We find 3 period oscillation in histogram of difference of consecutive primes also appears in r(n). Is there any relation between the sum and difference of prime number? The histogram of difference only contain the difference of consecutive primes, but the r(n) is the sum of any two primes. To make a clear comparison, we could also calculate the difference of two any primes before a number M.

Here we define:

$$D_M = \{d \mid d = p_i - p_j, i \neq j, p_{i,j} \in primes, p_{i,j} < M\};$$
$$D_M(n) = \| d \mid d = n, d \in D_M \|$$

$D_M(n)$ is the number of representations of an even number n as the difference of two primes before M. Here we select M=20000. The figure of series $D_M(n)$ and r(n) are plotted together (Fig 9).

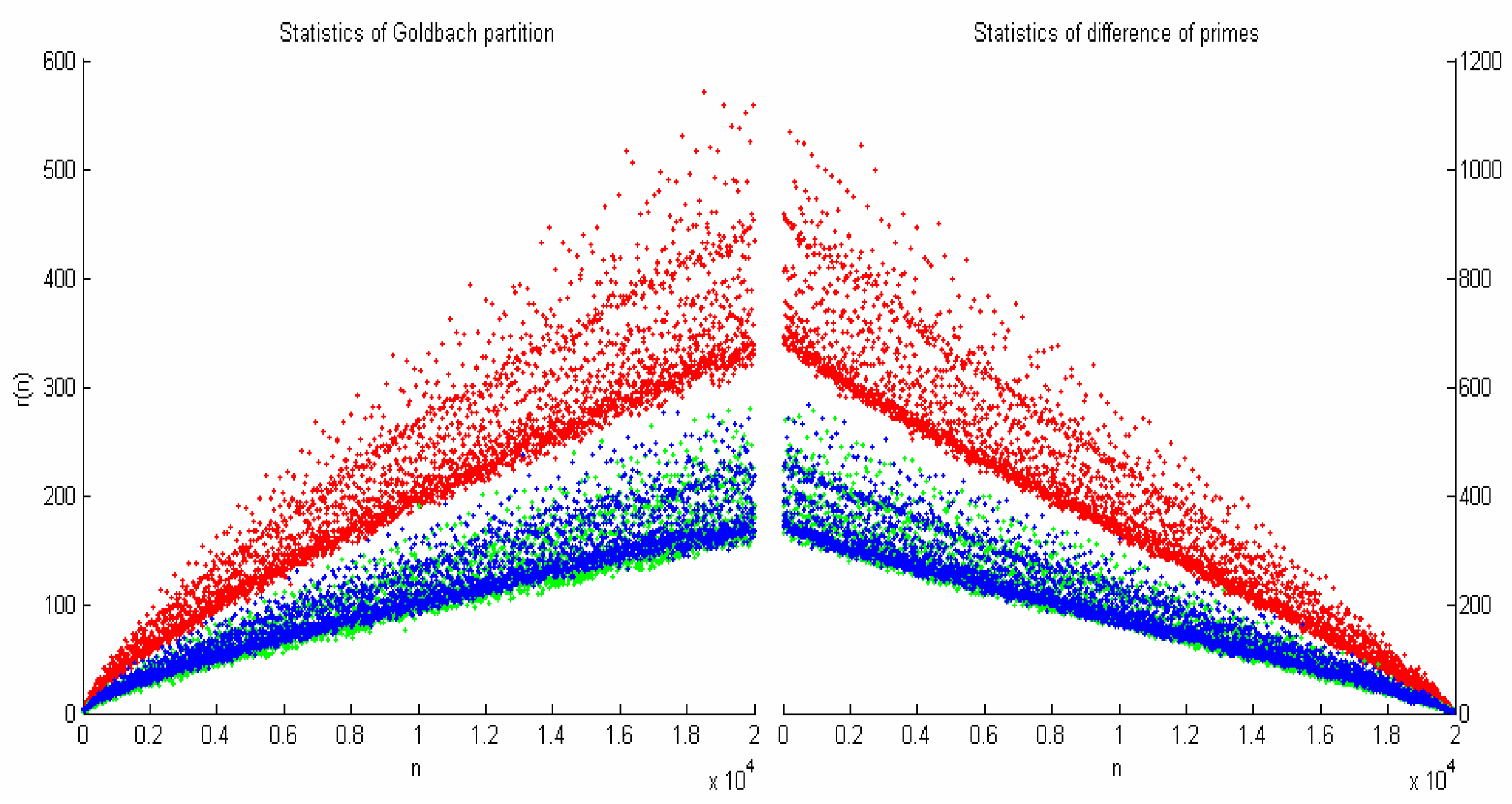

Fig 9. (a) Series r(n). (b) Series $D(n)$

We find D(n) is almost the mirror figure of r(n). The same fractal feature also appears in the statistics of difference of primes. 3 period oscillations in the histograms of difference of consecutive primes may be the special example of this phenomenon. If defining ‘3+97’ and ‘97+3’ as different representations:

$$R(n) = \begin{cases} 2r(n), n \notin primes \\ 2r(n) - 1, n \in primes \end{cases}$$

We get another conjecture:

**Conjecture 3:**

**When $n \to \infty$, the figure of series R(n) and D(n) will be precise symmetry.**

It has no surprise because the conjecture of Hardy-Littlewood for D(n) and R(n) are similar expression. We believe this symmetry will play an important role in solving some open problems in number theory. For example, if we admit the increases of r(n) and this symmetry, the Twin Primes conjecture will be true. This hypothesis states that there are infinite pairs of consecutive primes with difference of 2.

This is a little bit confusing to understand that h(n) has the precisely same detailed structure as r(n) . We guess Hardy-Littlewood may know these phenomena clearly, but there was no computer to buy in 1923. Here we just give a clearer depiction and a fashionable name “fractal” for them. .Now there is only few research papers for the fractal in primes system. Most of them could be found in web site [18]. Wish this simple and obvious fractal could spark some new ideas in this area.

## 3 Theory analyzing

Several current researches are mainly for the lower bounds of r(n). Here we are concerned with the periodic behavior of r(n).The distribution of the prime numbers among the integers seems somewhat random. So the combination of two prime sequences may be more complex. But from the viewpoint of chaos and fractal, the most complex phenomena are always produced by simple law. Here we just give a simple description for the dynamical character of r(n) but makes no attempt at explaining it.

Series r(n) is composed by different period series. If we only care its periodic character and don’t consider their value, this period embedding and compositing operation could be easily described by the “*” product in symbolic dynamics theory. Using this theory, we could also connect series r(n) with some one dimension unimodal mappings.

Here we give a brief introduction for “Applied symbolic dynamics”[19]. Symbolic dynamic is a profound pure math theory, but its application in chaos analyzing, “Applied symbolic dynamics” has developed into a simple and powerful tools for researchers in nonlinear science. Symbolic dynamics is a coarse-grained description of dynamics by taking into account the “geometry”. It could be easily explained by one-dimensional unimodal mappings $x_{n+1} = f(u, x_n)$ (Fig 10).

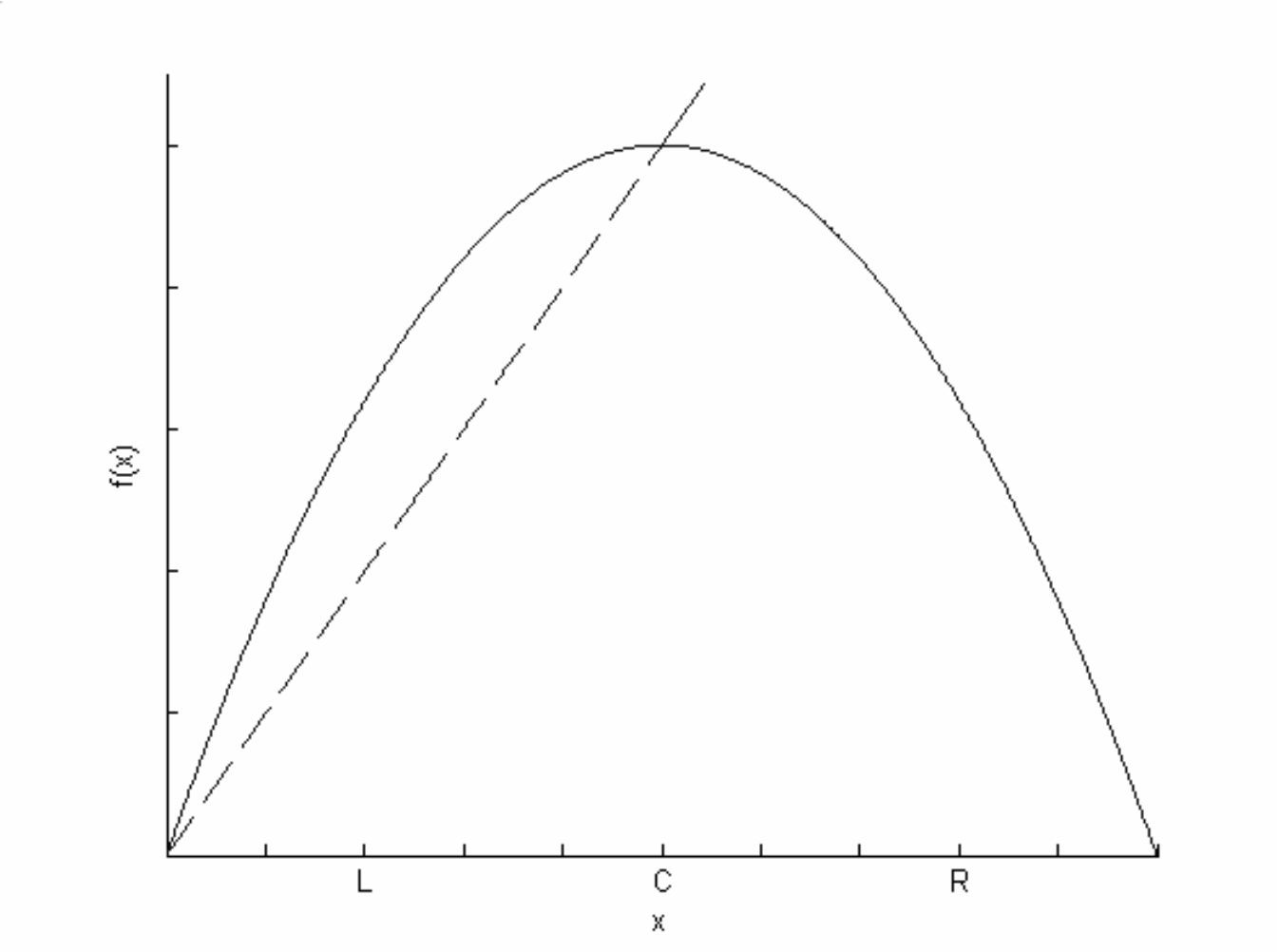

Fig 10. Symbol sequence in one dimension unimodal mapping

In Fig6, point ‘c’ in x axis corresponds to the maximal values of function $f(u, x)$. It divides the x axis into left part and right part. Through the iterated function $x_{n+1} = f(u, x_n)$ we could get an orbit from an initial point $x_0$: $x_0, x_1 = f(x_0), x_2 = f(x_1), \cdots\cdots, x_n = f(x_{n-1})\cdots\cdots$

If we only care the relative place of $x_i$: the point on the left of 'c' is marked as 'L' and the right is 'R. So this orbit could be described by a symbolic sequence like $'LLLRRRC\cdots', here, x_i = \begin{cases} L, x_i < c \\ C, x_i = c \\ R, x_i > c \end{cases}$.

Normally, we use basic sequence to represent the whole period sequence, for example $RL \leftrightarrow RLRLRL\cdots\cdots$

Our discussion mainly refers to three rules of symbolic dynamics:

(1) Kneading symbolic sequence. Any orbit can be described by a symbolic sequence, but some symbolic sequences couldn't correspond to an actual orbit. MSS table describe the basic 'admissible' sequence. Such sequence is also called kneading symbolic sequence, which describe the orbits beginning from the maximal value point of mapping.

(2) "*" product. It defines a composition rules that generate more admissible sequences from known ones, which is called DGP rule.

Here is a limited length sequence $P$ and sequence $Q = q_1 q_2 \cdots$

$$P * Q = (P * q_1)(P * q_2)\cdots,$$

$$P * q_i = \begin{cases} (PC)_+, q_i = R; \\ PC, q_i = C; \\ (PC)_-, q_i = L. \end{cases}$$

$$(PC)_+ = \max\{PR, PL\}, (PC)_- = \min\{PR, PL\}.$$

(3) The relation between symbolic sequence and parameter in a mapping. For a special mapping $x_{n+1} = f(u, x_n)$, we could get a parameter $u$ from one kneading symbolic sequence. That means we could get the actual orbit from a kneading sequence.

To describe the period oscillations of r(n), we select some primes number length of kneading symbolic sequences:

$$M_3 = RLC;$$
$$M_5 = RLLLC;$$
$$\vdots$$
$$M_{pi} = RL\cdots C;$$

$p_i$ is ith prime number.

For example, $M_3$ describes the 3 period oscillations. $M_3 * M_5$ is 3*5=15 period. It describes the period oscillations which is 3 periods in coarse-grained description, and these three parts are all comprise of 5 period oscillations. Obviously, the periodic oscillations behavior of series $r(n), n \to \infty$ could be described by the symbolic sequence:

$M_r = M_3 * M_5 * M_7 * \cdots * M_{pi} * \cdots$, $p_i$ is ith prime number.

Because all the $M_{pi}$ are kneading sequence, their '*' composition is also kneading sequence. In theory, there must be an actual orbit in a certain mapping $x_{n+1} = f(u, x_n)$ corresponding to this sequence. This orbit will have same periodic behavior as r(n). Through symbol sequence, we could build the relation between r(n) and some simple one dimension mappings.

Here we select: $f(x,u) = 1 - ux^2, x \in [-1,1], u \in (0,2]$. According to rule (3), we could get a parameter $u$ corresponding to $M_r$.We build a 3 period series first. For $M_3 = RLC$, we could get its corresponding parameter $u(M_3) \approx 1.754877666$. The orbit of $x_{n+1} = 1 - 1.754877666 x_n^2, x_0 = 1$ is shown in (Fig 11.a).

Then for $M_3 * M_5 = RLC * RLLLC = RLLRLRRLRRLRRLC$ , We could get $u(M_3 * M_5) \approx 1.79002267$. The orbit of $x_{n+1} = 1 - 1.79002267 x_n^2, x_0 = 1$ is shown in Fig 11.b. A segment of r(n) is also shown in Fig 11 c,d.( Fig 11.d separates the two lower parts)

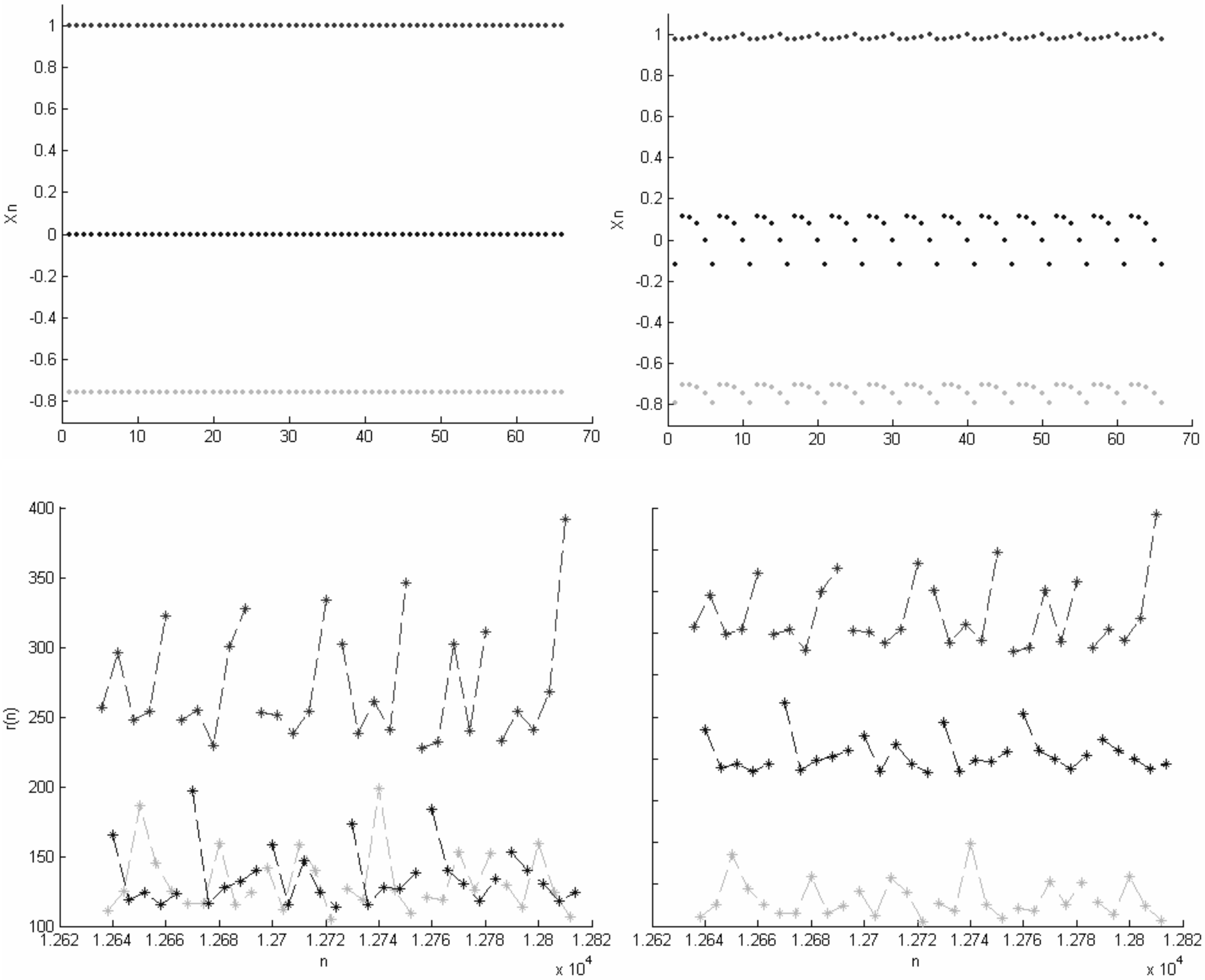

Fig 11. (a) The orbit of $x_{n+1} = 1 - 1.754877666 x_n^2, x_0 = 1$. (b) $x_{n+1} = 1 - 1.79002267 x_n^2, x_0 = 1$.(c),(d) a

segment of r(n)

Just as we supposed, Fig 8.a is 3 period and Fig 8.b has three parts and every part is 5 period oscillations. Because $M_3 * M_5 * M_7$ is a 105 length symbolic sequence, the calculation of $u(M_3 * M_5 * M_7)$ will become very difficult for precision limitation. Here we just want to show the complex dynamic character of series r(n) could be presented by a simple iterated function.

$u(M_r)$ could be estimated as follows:

Here we select $M_5 = RLLLC$

$$M_3 * M_5 = RLLRLRRLRRLRRLC;$$

Because the first two symbols of kneading sequence are '$RL$', so:

$$M_3 * M_5 * M_7 = RLLRLRRLRRLRRLC * RL \cdots C$$
$$= (RLLRLRRLRRLRRLR)(RLLRLRRLRRLRRLL) \cdots C$$

According to the * composition rule, the $M_r$ must have the form:

$$M_r = (RLLRLRRLRRLRRLR)(RLLRLRRLRRLRRLL) \cdots C = \sum LL \cdots C;$$
$$here, \sum = (RLLRLRRLRRLRRLR)RLLRLRRLRRLRR.$$

So:

$$\sum RC < M_r < \sum LC;$$
$$u(\sum RC) < u(M_r) < u(\sum LC).$$

We could get:

$$u(\sum RC) = 1.7900227173230200961029368045274;$$
$$u(\sum LC) = 1.7900240852603652719210458599264;$$
$$1.7900227 < u(M_r) < 1.7900241.$$

In theory, $x_{n+1} = 1 - u(M_r)x_n^2$ could describe the periodic character of series r(n). Here we can't prove they are topology conjugated, but the a referenced Lyapunov exponent of r(n) could be obtained through this connection. The Lyapunov exponent of $x_{n+1} = 1 - ux_n^2, 1.7900227 < u < 1.7900241$ is shown in Fig 12.

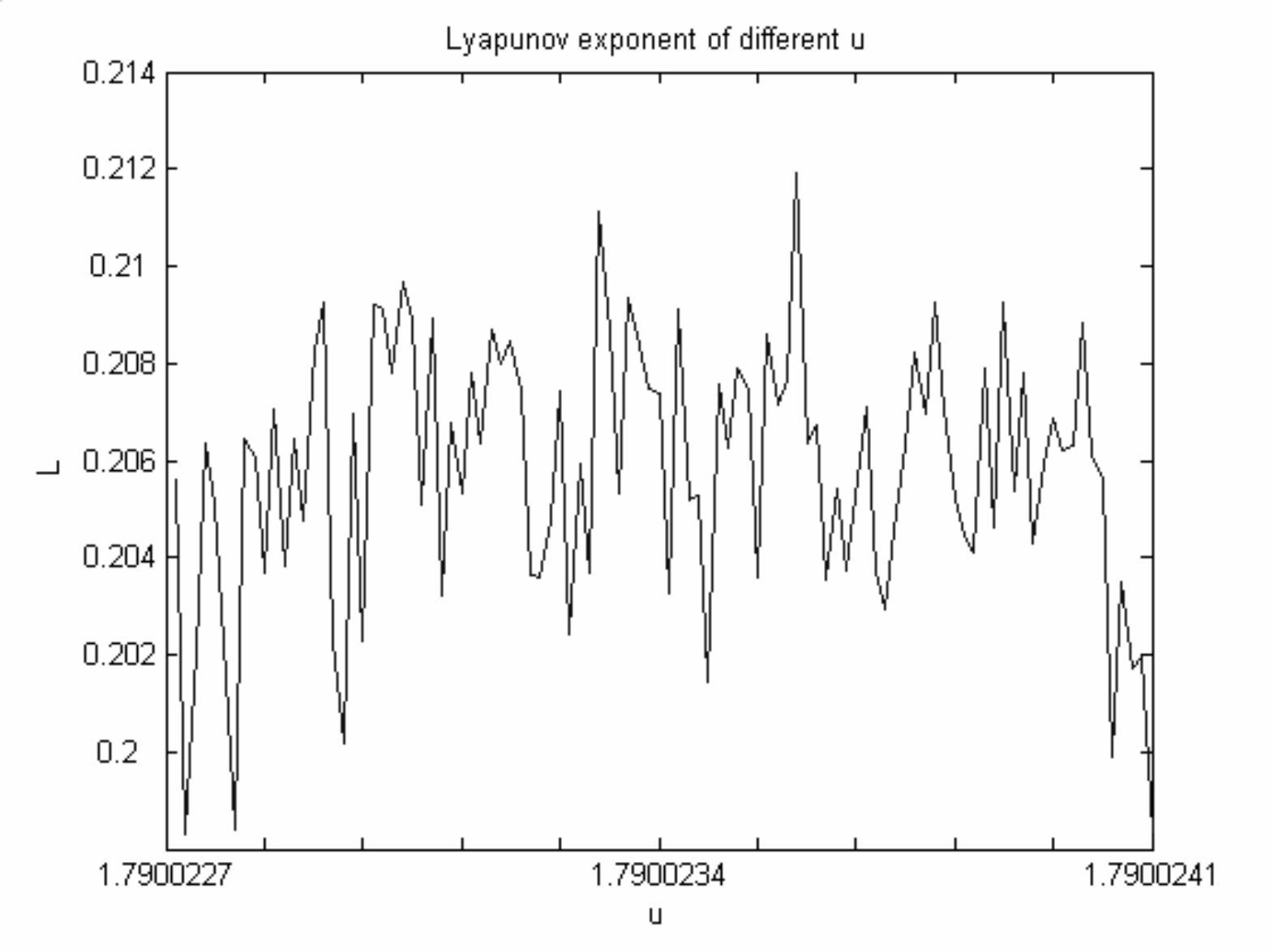

Fig 12. Lyapunov exponent of $x_{n+1} = 1 - ux_n^2, 1.7900227 < u < 1.7900241$

So under this meaning, Lyapunov exponent of series $r(n), n \to \infty : L(r) \in (0.1983, 0.2119)$. There are two other expressions for $M_5$, we could also estimated the related $u$ and Lyapunov exponent:

For $M_5 = RLLRC$, $1.7877578 < u < 1.7879103$, $L(r) \in (0.1163, 0.1850)$.

For $M_5 = RLRRC$, $1.7836878 < u < 1.7844579$, $L(r) \in (0.0562, 0.1590)$

In these three zones of u, we could find some orbits of $x_{n+1} = 1 - ux_n^2$ that have the same periodic oscillation behavior as series r(n). Because they all have the positive Lyapunov exponent, we could say the r(n) is a chaos series. In fact, because the value zone of series r(n) is not constant , it's also very difficulty to estimate the real Lyapunov exponent by chaos time series analyzing technology.

Goldbach Conjecture connects a multiplicative property—being prime—to addition. Such connections are poorly understood today. If we regard the primes as prime number period oscillations, the multiplicative property and addition property of primes could be connected by '*' product in symbolic dynamics. It may provide a possible method for success in cracking the Goldbach Conjecture. Renormalization group theory is another powerful tool to study the fractal and chaos. But to analyze this fractal, we may need the deep knowledge in number theory and physics. Some useful links about this cross-research could be found in web site [20].

## 4 Conclusions

Just like many of these researches, we could only give some interesting experimental results and a rough explanation. Prime numbers appear to us as a random collection of numbers without any structure, but their sum or difference turns into the well-regulated fractal. Here we cite a word of Y. Motohashi[21] to describe the situation of Riemann Hypothesis and many other open problems in number theory, "it will be settled without any fundamental changes in our mathematical thoughts, namely, all tools are ready to attack it but just a penetrating idea is missing." Could this “Goldbach Comet" knock a hole in the door of prime treasure house?